\documentstyle[11pt]{article}

\newfont{\msbm}{msbm10}

\def\gg{\hbox{\tt g}}
\def\g{\gamma}

\def\CC{{\cal C}}

\def\L{{\cal L}}

\def\P{{\cal P}}

\def\A{{\cal A}}

\def\R{\hbox{\msbm R}}

\def\C{\hbox{\msbm C}}

\def\H{{\cal H}}

\def\m{\mu}

\def\om{\mbox{$\omega$}}

\def\s{\sigma}

\newtheorem{lem}{Lemma}
\newtheorem{prop}{Proposition}
\newtheorem{cor}{Corolary}
\newtheorem{theo}{Theorem}
\newtheorem{deff}{Definition}

\def\bc{\begin{cor}}
\def\ec{\end{cor}}

\def\bt{\begin{theo}}
\def\et{\end{theo}}

\def\bd{\begin{deff}}
\def\ed{\end{deff}}

\def\bp{\begin{prop}}
\def\ep{\end{prop}}

\def\ba{\begin{eqnarray}}
\def\ea{\end{eqnarray}}

\def\be{\begin{equation}}
\def\ee{\end{equation}}

\newfont{\msbms}{msbm6}  

\def\E{{\cal E}}

\def\G{{\cal G}}

\def\mo{\mbox{$\mu_0$}}
\def\mol{\mbox{${\mu_0}_L$}}

\def\Sig{\Sigma}

\def\Ab{\mbox{$\bar A$}}
\def\ab{\mbox{$\bar \A$}}
\def\Si{\mbox{$\Sigma$}}

\def\rep{representation}
\def\reps{representations}

\begin{document}


\title{On the structure of the space of \\generalized connections}

\author{J. M. Velhinho}

\date{{\it Departamento de F\'\i sica, Universidade da Beira 
Interior}
\\R. Marqu\^es D'\'Avila e Bolama,
6201-001 Covilh\~a, Portugal\\jvelhi@dfisica.ubi.pt}

\maketitle


\begin{abstract}

\noindent  
We give a modern account of the construction and structure of the space of 
generalized connections, an extension of the space of connections that plays a central
role in loop quantum gravity.
\end{abstract}

\tableofcontents
\pagestyle{myheadings}


\section{Introduction}
\label{int}
Loop quantum gravity is an attempt to canonically quantize general relativity,
starting from a $SU(2)$ gauge formulation of the classical theory,
and using non-perturbative background independent methods.

The purpose of the present review is to give a  modern and updated 
account of the
structure of the so-called space of generalized (or distributional) connections \ab, an
extension of the classical space of smooth connections $\A$, that plays a central role 
in loop quantum gravity.
Since analogous spaces of generalized connections can be constructed for  different
compact gauge  groups, this discussion may also be of interest for other (especially
diffeomorphism invariant)   models.

Although we will certainly not attempt to make a survey of loop quantum 
gravity\footnote{For reviews of loop quantum gravity, or of its by now numerous and very 
impressive achievements and applications, we recommend
the excelent available works, ranging from quick surveys to very thorough and 
comprehensive treatments, by some of the most active researchers in the 
field~\cite{A1,A2,A3,R1,R2,Sm,T1,T2,ALR}.},
a minimal introduction to the general programme and its foundations is required, in order 
to appreciate the origins, place and role
of the space of generalized connections.
This is the purpose of the current section.

The starting point for loop quantum gravity is the $SU(2)$ version of Ashtekar's 
canonical formulation
of   general relativity as a special kind of   gauge theory~\cite{A1,A4,Ba}. The phase space 
$\A\times\E$ is made of
canonically conjugate pairs $(A,E)$, where $A\in\A$ are smooth connections on the spatial
manifold $\Sigma$ and $E\in\E$ are electric fields, i.e.~$su(2)^*$-valued vector densities of 
weight one. As is well known, the theory has no Hamiltonian (in the usual sense),
only constraints.  Besides the Gauss constraint, that generates $SU(2)$ gauge 
transformations, there is the (spatial) diffeomorphism constraint, that generates
spatial diffeomorphisms, and the Hamiltonian constraint, associated with 
reparametrizations of time.

Loop quantum gravity follows the Dirac method for the quantization of theories with 
constraints. In this method, one  tries to impose the constraints after quantization of the
unconstrained phase space. A kinematical algebra, i.e.~a Poisson algebra of functions 
that separate points in the unconstrained phase space, must therefore be chosen, 
as the set of elementary variables to be quantized. The specific choice of variables is one
of the most characteristic aspects of loop quantum gravity, and the one that gave it
its name. It was introduced in quantum gravity by Jacobson, Rovelli and Smolin~\cite{JS,RS}.
In its modern formulation, the loop quantum gravity kinematical
variables are: {\it i}\/) continuous complex functions of holonomies $A\mapsto h_c(A)$ of 
connections along certain curves $c$, as configuration variables, and {\it ii}\/)
electric flux functions $E\mapsto E_{S,f}(E):=\int_S \star E_j f^j$ on surfaces $S$,
where $(\star E_j)_{\m\nu}=E^{\alpha}_j\epsilon_{\alpha\m\nu}$ and $f$ are 
$su(2)$-valued functions, as momentum variables\footnote{There are, in fact, subtleties
in the definition of the Poisson algebra generated by these functions, especially concerning 
the flux variables. A satisfatory answer to this issue was given in~\cite{ACZ}.}.

Besides the non-trivial fact that a well defined quantization of such an algebra 
was constructed~\cite{AI,AL1,AL3,ALMMT}, these variables have in their favor the fact that 
they are well adapted
to the constraints. Notice that, replacing the crucial role of the Hamiltonian
in standard quantum field theory, the constraints are now the fundamental guidelines
in the construction/selection of the quantization, as they must be implemented,
either as self-adjoint operators or as unitary \reps\ of 
the corresponding groups. It is therefore welcomed to start with classical variables that have 
a simple 
behaviour with respect to the constraints. In this respect,
notice that under a $SU(2)$ gauge transformation $\gg$,
the flux variables transform among themselves, $E_{S,f}\mapsto E_{S,\gg^{-1}f\gg}$,
and holonomies transform simply as $h_c\mapsto \gg(b)h_c\gg^{-1}(a)$,
where $a$ and $b$ are the starting and end points of the curve $c$. Moreover, gauge
invariant configuration functions are easily obtained, by considering closed curves (loops) $c$
and taking the trace. In what concerns diffeomorphisms, the important fact is that holonomies and 
flux variables are intrinsically,
background independently, defined integrals, since, locally, connections $A$ are 1-forms and 
the objects
$\star E_jf^j$ are 2-forms.  Simple covariance properties then follow:
under a spatial diffeomorphism $\varphi$ we have
$h_c\mapsto h_{\varphi^{-1}c}$ and $E_{S,f}\mapsto E_{\varphi^{-1}S,\varphi*f}$.
As for the Hamiltonian constraint, early formal arguments~\cite{JS,RS} suggested that
a quantum version of it could be defined within this framework, and that certain 
"loop states" could be solutions of the "quantum Hamiltonian constraint". 

The space of generalized connections \ab\ is an extension of $\A$
determined by the above
configuration variables, in the sense that those variables are quantized as functions 
on \ab.
Moreover,  the flux variables are naturally realized as derivations of a certain algebra of 
functions in \ab. 
Thus, \ab\ plays the role of a
"universal quantum configuration space", i.e.~the space where a Schrodinger-like $L^2$ \rep\
(of the chosen variables) is naturally defined. 
Notice that this situation is reminiscent
of the well known  quantization of scalar fields, where a distributional extension
of the space of classical smooth fields is required in order that measures and 
corresponding $L^2$ spaces can
be defined. In the present case, we are led to a compact Hausdorff space \ab,
and (regular Borel) measures are thus guaranteed to exist. 

Before we go into any details, let us fix the particular framework considered
in the present review.

There are several versions of the space of generalized connections, depending
on the differentiability class of the curves $c$ used in the holonomies.  We will
assume that the spatial manifold $\Sigma$ is endowed with a (real)
analytic structure and that the curves $c$ are piecewise analytic (analytic surfaces are
accordingly used in the flux variables). While this may seem  unnatural,
there are reasons to believe that the analytic set-up is sufficiently general.  
As it avoids technical complications arising in more general settings, the analytic case is
the most studied, and the one in which more rigorous results have been obtained. 
Nevertheless, important
progress has been made in the case of piecewise smooth, or more general, 
curves~\cite{SB1,SB2,LT,Fl}.

Notice also  that we will describe the non gauge invariant space 
of generalized 
connections, as opposed to the gauge invariant space ${\overline {\A/\G}}$ of generalized 
connections modulo
gauge transformations, in which loop variables defined by closed curves are
used. This was in fact the original approach~\cite{AI,AL1}, but nowadays the non gauge 
invariant space
\ab~\cite{B1,AL3} is typically preferred, leaving the solution of the Gauss constraint to a 
later stage.
The relation between \ab\ and ${\overline {\A/\G}}$, or how to solve the Gauss constraint
in the \ab\ framework, is very well understood; the two approaches are seen to be fully 
equivalent.

Let us then see how the  space of generalized connections appears in loop quantum gravity.
The crucial fact is that the chosen set of configuration variables
\be
\label{I1}
A\mapsto F(h_{c_1}(A),\ldots,h_{c_n}(A)),\ F\in C((SU(2)^n),
\ee
is a unital $\star$-algebra of bounded functions, due to the 
compactness of the gauge group (this is called the algebra  of cylindrical 
functions ${\rm Cyl}(\A)$). The natural sup norm $C^{\star}$-completion of this algebra, 
introduced
by Ashtekar and Isham~\cite{AI}, is called the holonomy algebra $\overline {\rm Cyl}(\A)$.
One then requires that the quantization of kinematical variables 
produces a 
representation of the unital commutative $C^{\star}$-algebra 
$\overline {\rm Cyl}(\A)$.

By the Gelfand--Naimark fundamental characterization of commutative $C^{\star}$-algebras, we
know that every commutative unital $C^{\star}$-algebra is (isomorphic to) the algebra $C(X)$ 
of continuous complex functions on a unique (up to homeomorphism) compact Hausdorff
space $X$. In the case of the holonomy algebra $\overline {\rm Cyl}(\A)$, 
the corresponding compact space
is naturally realized as the space of all morphisms from a certain
groupoid of paths (equivalence classes of curves) to the gauge group. This
is the space of generalized 
connections \ab. 

The relevance of \ab\ to the quantization process  now becomes
obvious. Again from general results, we know that: {\it i}\/) every representation
of $C(\ab)$ is a direct sum of cyclic representations;
{\it ii}\/) every  cyclic \rep\ is  a \rep\ by multiplication operators on
a Hilbert space $L^2(\ab,\m)$, where $\m$ is a measure on \ab. Thus, given the isomorphism
$\overline {\rm Cyl}(\A)\cong C(\ab)$, every quantization 
of the loop quantum gravity kinematical algebra decomposes into a direct sum
\be
\label{I2}
\H=\bigoplus_i \H_i=\bigoplus_i L^2(\ab,\m_i),
\ee
with $\overline {\rm Cyl}(\A)$ being  represented in each space $L^2(\ab,\m_i)$ by multiplication
operators:
\be
\label{I3}
({\widehat f}\psi)(\bar A)=\check{f}(\bar A)\psi(\bar A),
\ee
where $\psi\in L^2(\ab,\m_i)$, $f\in\overline {\rm Cyl}(\A)$ and 
$\check{f}\in C(\ab)$ is the image of $f$. 

The study of the structure of \ab\ (and $\overline {\A/\G}$) and of measures thereon 
was done in the first half of the 1990s~\cite{AL1,B1,B2,MM,AL2,AL3}. 
After the seminal work of Ashtekar and Isham~\cite{AI}, Ashtekar and Lewandowski
introduced the analytic framework and succeeded in the construction of a very natural measure,
distinguished by its
simplicity and invariance properties~\cite{AL1}. This is the so-called Ashtekar--Lewandowski, 
or uniform, measure \mo. Although
several other measures were constructed, it turns out that the uniform  measure still stands 
as the only known measure that gives rise to a \rep\ of the full kinematical algebra
of holonomies and flux variables. In other words, the corresponding {\it kinematical}\/ 
Hilbert space
$\H_0:=L^2(\ab,\mo)$ supports the only known \rep\ of the kinematical 
algebra of loop quantum gravity~\cite{ALMMT}.

This $\H_0$ \rep\ is cyclic with respect to $C(\ab)$, 
of course, and is irreducible, as it should be, under the action of the kinematical 
algebra~\cite{ST1}. Moreover, the measure \mo\ is gauge invariant and invariant under
the action of analytic diffeomorphisms. This immediately leads to the required 
unitary \reps\ of both groups. Thus, as far as the Gauss and diffeomorphism 
constraints are concerned, the $\H_0$ \rep\ seems to qualify as an intermediate, or auxiliary,
\rep\ of kinematical variables and constraints, as required  in the Dirac method. 
The above facts alone
are sufficient to justify the absolutely central role of the $\H_0$ \rep\ in the
canonical loop quantum gravity programme.
The $\H_0$ \rep\ 
is the kinematical \rep\ used in loop quantum gravity;
virtually  all further developments are based upon it.

The seemingly unique status of the $\H_0$ \rep\ was recently reinforced by a detailed analysis
of the \rep\ theory of the kinematical algebra~\cite{S1,S2,OL,ST}. Although the uniqueness
of the $\H_0$ \rep\ was not fully established, it was shown~\cite{OL,ST} that an {\it a priori}\/ large
class of \reps, that also support a unitary implementation
of the group of analytic diffeormophisms, contains in fact only reducible \reps,
and that every irreducible component is equivalent to the $\H_0$ \rep.

Finally, for completeness, 
let us stress that the $\H_0$ \rep\ is not,
by far, the end of the quantization process. Important as it is, $\H_0$ is the starting
point for the hardest and most interesting part of the quantization, and this is precisely
the reason why it is so important that $\H_0$, and therefore \ab, are well defined and 
well understood. 

Once the constraints are represented, one must, of course, solve them.
As already mentioned, the Gauss constraint is easily dealt with. It can be solved before
or after solving the other constraints. If we choose to solve it prior to the other
constraints, the solution is the (large) subspace of gauge invariant 
elements of $\H_0$.   

On the contrary, already the diffeomorphism constraint cannot be solved within $\H_0$.
In fact, the necessity of distributional, or generalized, solutions is typical of the
Dirac method, when non-compact invariance groups, as the diffeomorphism group, are 
involved. Starting from $\H_0$, a complete space of solutions of the diffeomorphism 
constraint was constructed in a 
distributional extension of $\H_0$, and equipped with an inner product induced by that of 
$\H_0$~\cite{ALMMT}. 

All these efforts still leave ahead far more challenging tasks, like the construction
of the quantum Hamiltonian constraint and its space of solutions, recovery of (semi-)classical
physics, construction of observables, or interpretational issues, just to mention a few.
These subjects, and many more, are being actively pursued. Regarding e.g.~the crucial
and most difficult question of the Hamiltonian constraint, it is truly remarkable that
a candidate quantum operator has been rigorously defined in $\H_0$~\cite{T1}. Although there are
open questions regarding the physical correctness of this operator, 
the fact that all constraints
can be implemented  reinforces, again, the status of $\H_0$, and of the non-perturbative and 
background independent methods used in loop quantum gravity. More recently, a new 
proposal~\cite{T3} suggests the possibility of defining  the Hamiltonian
constraint directly on the space of solutions of the diffeomorphism constraint. If the expectations raised by that 
work are confirmed, then important technical as well as conceptual simplifications may
occur, possibly leading to further and quicker progress.

Let us conclude with an overview of the present work.
In section 2 we present the space of generalized connections \ab\  
and discuss its important projective structure. As first pointed out by Baez~\cite{B3}
and later on explicitly put forward in~\cite{Ve}, the proper framework to express the algebraic 
properties of \ab\ implicit in earlier formulations~\cite{B1,AL3} uses the language of category
theory. The space \ab\ is actually a space of functors, or morphisms.
(The notions from category theory  used in the present work are minimal
and elementary -- we review them in section \ref{cat}.)
The algebraic properties of \ab\ reflect simply the algebraic properties of parallel transports,
as functions of curves. These functions  depend only on certain equivalence classes
of curves, and this is precisely how the groupoid of paths $\P$, discussed in section
\ref{seg}, emerges. As a bare set, \ab\ is the set of functors from $\P$ to the gauge group $G$.  
This large set is actually a limit of a family of finite dimensional spaces, each of which
is identified with some power $G^n$ of the gauge group. It is by means of this characterization
that \ab\ is turned into an interesting and manageable space, with a rich 
structure~\cite{AL1,AL2,AL3,MM}. This so-called projective structure is discussed in some detail
section \ref{pro}, where we concentrate on topological aspects. Section \ref{inductive} 
describes the "dual"
inductive structure of the algebra of functions on \ab.
In section \ref{gauge} we show how gauge transformations and diffeomorphisms fit within the category
formulation of \ab, and speculate on possible extensions of the diffeomorphism group
suggested by this formulation.

After the description of \ab, we return in section \ref{new3}
to its physical role as a possible "quantum configuration space" for theories of connections.
In section \ref{classical} we display \ab\ as the compact space defined by the holonomy algebra.
Section \ref{mea} is dedicated to the general structure of measures on \ab.
Finally, we  present the Ashtekar-Lewandowski measure \mo\ and briefly discuss the most important
properties of the corresponding \rep.

\section{The space of generalized connections $\bar \A$}
\label{sec2}
\subsection{Elementary notions from category theory}
\label{cat}
A (small) category $\tt C$ is formed by "arrows" (or "morphisms") between
"objects". The set of objects is denoted by ${\rm Obj}\,{\tt C}$ and the set
of arrows by ${\rm Mor}\,{\tt C}$ (or simply by ${\tt C}$). 
For $x,y\in{\rm Obj}\,{\tt C}$, ${\rm Hom}_{\,\tt C}\,[x,y]$ denotes the set
of all arrows from $x$ to $y$. The following two maps 
$r,s:{\rm Mor}\,{\tt C}\to{\rm Obj}\,{\tt C}$, called range and 
source, respectively, are naturally defined: $r(\g)$ is the object on
which the arrow $\g$ ends, i.e.~$r(\g)=x$ if 
$\g\in{\rm Hom}_{\,\tt C}\,[\,\cdot,x]$; likewise, $s(\g)$ is the object on 
which $\g$ starts. The main characteristic of a category is the existence
of an associative composition operation between compatible arrows, meaning
that there are binary operations
\be
\label{1}
{\rm Hom}_{\,\tt C}\,[x,y]\times {\rm Hom}_{\,\tt C}\,[y,z]\to 
{\rm Hom}_{\,\tt C}\,[x,z],\ \ (\g,\g')\mapsto \g'\g
\ee
satisfying $\g''(\g'\g)=(\g''\g')\g$.
It is also required that for every object $x$ there exists a unique identity 
arrow ${\bf 1}_x\in{\rm Hom}_{\,\tt C}\,[x,x]$, such that 
${\bf 1}_x\gamma=\gamma$, $\forall\gamma
\in{\rm Hom}_{\,\tt C}\,[\,\cdot,x]$ and $\gamma{\bf 1}_x=\gamma$, 
$\forall\gamma
\in{\rm Hom}_{\,\tt C}\,[x,\cdot\,]$.
Naturally, an arrow 
$\gamma\in{\rm Hom}_{\,\tt C}\,[x,y]$ is said to be invertible if there exists 
$\gamma^{-1}\in{\rm Hom}_{\,\tt C}\,[y,x]$ such that $\gamma^{-1}
\gamma={\bf 1}_x$ and $\gamma\gamma^{-1}={\bf 1}_y$.

Most important to our discussion is the notion of groupoid, which is simply
a category in which every arrow is invertible.

Groups are a special class of groupoids, and therefore of categories.
In this case the arrows are the elements of the group, composed through
the group operation. As a category, a group has a single object, namely
the group identity.

A map between categories that preserves the algebraic structure is called
a functor. A functor $F:{\tt A}\to {\tt B}$ is  made of two maps
(usually denoted by the same symbol), between objects and between arrows,
subjected to the conditions $F({\bf 1}_x)={\bf 1}_{F(x)}$,
$\g\in{\rm Hom}_{\,\tt A}\,[x,y]$ implies
$F(\g)\in{\rm Hom}_{\,\tt B}\,[F(x),F(y)]$ and $F(\g'\g)=F(\g')F(\g)$.

Another important notion is that of a natural transformation between
functors. For two functors $S,T: {\tt A}\to {\tt B}$, a natural transformation
$\tau:S\to T$ is a map from ${\rm Obj}\,{\tt A}$ to ${\rm Mor}\,{\tt B}$,
assigning to each $x\in {\rm Obj}\,{\tt A}$ an arrow $\tau(x)\in 
{\rm Hom}_{\,\tt B}\,[S(x),T(x)]$ such that $T(\g)\tau(s(\g))=\tau(r(\g))S(\g)$,
$\forall \g\in {\rm Mor}\,{\tt A}$. If, in particular, ${\tt B}$ is a group,
one sees that natural transformations give us a representation of the 
product group
$\times_{x\in {\rm Obj}\,{\tt A}}{\tt B}$, acting on the set of all
functors $F:{\tt A}\to {\tt B}$, by
\be
\label{2}
F\mapsto \tau F:\ \ \tau F(\g)=\tau(r(\g))F(\g)\tau(s(\g))^{-1}.
\ee
\subsection{The groupoid of paths $\P$}
\label{seg}
Let \Si\ be an analytic, connected, orientable and paracompact $d$-dimensional manifold. 
Let us consider the set $\CC$ of all continuous, oriented and piecewise analytic
parametrized curves in \Si , i.e.~maps
$$
c:[0,t_1]\cup\ldots\cup [t_{n-1},1]\to\Sig
$$
which are continuous in all the domain $[0,1]$, analytic in the closed
intervals $[t_k,t_{k+1}]$ and such that the images $c\bigl(]t_k,t_{k+1}[
\bigr)$ of the open intervals $]t_k,t_{k+1}[$ are submanifolds
embedded in \Si . In what follows, $\s(c):=c\bigl([0,1]\bigr)$ denotes the image of
the interval $[0,1]$. The maps $s$ (source) and $r$ (range) are
defined, respectively, by $s(c)=c(0)$, $r(c)=c(1)$. 

Given two curves
$c_1,c_2\in\CC$ such that $s(c_2)=r(c_1)$, let $c_2c_1\in\CC$
denote the natural composition given by
$$
(c_2c_1)(t)=\left\{\begin{array}{lll} c_1(2t), & {\rm for} & t\in[0,1/2] \\
c_2(2t-1), & {\rm for} & t\in[1/2,1]\,. \end{array} \right.
$$
This composition defines a binary operation in a well defined subset of
$\CC\times\CC$. Consider also the operation $c\mapsto c^{-1}$ 
given by $c^{-1}(t)=c(1-t)$. Notice that this composition
of parametrized curves is not truly associative, since the curves $(c_3c_2)c_1$ 
and $c_3(c_2c_1)$ are related by a reparametrization, i.e.~by an
orientation preserving piecewise analytic diffeomorphism $[0,1]\to [0,1]$.
Similarly, the curve $c^{-1}$ is not the inverse of the curve $c$.
We will refer to compositions of the form $c^{-1}c$ as retracings.
\begin{deff}
\label{def1}
Two curves $c,c'\in\CC$ are said to be equivalent if
\begin{itemize}
\item[{\rm (}i\/{\rm )}] $s(c)=s(c')\, ,\ r(c)=r(c')\, ;$
\item[{\rm (}ii\/{\rm )}] $c$ and $c'$ coincide up to a finite number of retracings
and a reparametrization.
\end{itemize}
\end{deff}
We will  denote the set of
all above defined equivalence classes by $\P$.
It is clear by ({\it i\/}) that the maps $s$ and $r$ are well
defined in $\P$. 
The image $\s$ can still be defined for special elements called edges. 
By edges we mean elements $e\in\P$ which are equivalence classes of
analytic (in all domain) curves $c:[0,1]\to\Si$. It is clear that the images
$c_1\bigl([0,1]\bigr)$ and $c_2\bigl([0,1]\bigr)$ corresponding to two equivalent analytic
curves coincide, and therefore we define $\s(e)$ as being $\s(c)$, where
$c$ is any analytic curve in the classe of the edge $e$.

We discuss next the natural groupoid structure on the set $\P$.
We will  refer to generic elements
of $\P$ as paths $p$, the symbol $e$ being reserved for edges.

The composition of paths
is defined by the composition of elements of $\CC$:  if $p,p'\in\P$ 
are such that $r(p)=s(p')$, one defines $p'p$ as the 
equivalence class of $c'c$, where $c$ (resp.~$c'$) belongs to the class
$p$ ($p'$). The independence of this composition with
respect to the
choice of representatives follows from condition ({\it ii\/}) above. The composition 
in $\P$ is now associative,
since $(c_3c_2)c_1$ and $c_3(c_2c_1)$ belong to the same equivalence class.

The points of $\Si$ play the role of objects in this context.
Points are in 1-1 correspondence with identity
paths:  given $x\in\Sig$ the corresponding identity ${\bf 1}_x\in\P$ is the
equivalence class of $c^{-1}c$, with $c\in\CC$ such that $s(c)=x$. 
If $p$ is 
the class of $c$ then $p^{-1}$ is the class of $c^{-1}$. It is clear that
$p^{-1}p={\bf 1}_{s(p)}$ and 
$p p^{-1}={\bf 1}_{r(p)}$. 

One, therefore, has a well defined 
groupoid, whose set of objects is $\Si$ and whose set of arrows is $\P$.
As usual, we will use the same notation, $\P$, both for
the set of arrows and for the groupoid.
Notice that
every element $p\in\P$ can be obtained as a composition of edges.
Therefore, the groupoid $\P$ is generated by the set of edges, although
it is not freely generated, since composition of edges may produce new edges.
\subsection{The set of functors ${\rm Hom}\,[\P,G]$}
\label{ssi}
Let $G$ be a (finite dimensional) connected and compact
Lie group.
\begin{deff}
\label{def01}
${\rm Hom}\,[\P,G]$ is the set of all 
functors from the groupoid $\P$ to the group $G$, i.e.~is the set of all
maps $\bar A:\P\to G$ such that $\bar A(p'p)=\bar A(p')\bar A(p)$ and
$\bar A(p^{-1})=\bar A(p)^{-1}$. 
\end{deff}
To be consistent with the literature, we will refer to elements of 
${\rm Hom}\,[\P,G]$ not as functors, but as morphisms, or  as
generalized (or distributional) connections, for reasons to be discussed next.

Let us 
show that the space $\A$ of smooth $G$-connections on any given principal 
$G$-bundle over $\Si$ is realized as a subspace of ${\rm Hom}\,[\P,G]$.
We think of this bundle as being associated to a classical field theory of connections,
and so we will also refer to $\A$ as the classical configuration space.
We will assume that
a fixed trivialization of the bundle  has been chosen. Connections
can then be identified with local connection potentials.

The space of connections $\A$ is injectively mapped into ${\rm Hom}\,[\P,G]$ through the
use of the parallel transport, or holonomy, functions. The holonomy
defined by a  connection $A\in\A$ and a  curve $c\in\CC$
is denoted by $h_c(A)$. Using the fixed trivialization, one can assume that
holonomies $h_c(A)$ take values on the group $G$. The following properties
of holonomies are seen to hold: 
{\it i}\/) $h_c(A)$ is invariant under 
reparametrizations of $c$; {\it ii}\/) $h_{c_1 c_2}(A)=h_{c_1}(A)h_{c_2}(A)$ 
and 
{\it iii}\/) $h_{c^{-1}}(A)={h_c(A)}^{-1}$. One thus conclude that 
$h_c(A)$ depends only
on the equivalence class of $c$, i.e., for fixed $A\in\A$, $h_c(A)$ defines a function
on $\P$, with values on $G$. This function is, moreover, a morphism, by {\it ii}\/).
Summarizing, we have a map $\A\to {\rm Hom}\,[\P,G]$:
\be
\label{image}
A\mapsto \bar A_A,\ \bar A_A(p):=h_p(A),\ \forall p\in\P.
\ee
That this map is injective is guaranteed by the crucial fact that the set
of holonomy functions $\{h_c,\ c\in\CC\}$ separates points in 
$\A$~\cite{G},
i.e.~given $A\not =A'$ one can find $c\in\CC$ such that $h_c(A)\not = h_c(A')$,
which is, of course, the expression of injectivity. The classical space $\A$
can then be seen as a subspace of ${\rm Hom}\,[\P,G]$.

The set ${\rm Hom}\,[\P,G]$ is, however, larger than the classical space $\A$.
To begin with, depending on $\Si$ and $G$, different bundles may exist, and
${\rm Hom}\,[\P,G]$ contains the space of connections of all these bundles.
Moreover, elements of ${\rm Hom}\,[\P,G]$ that do not correspond 
to any smooth connection do exist (see e.g.~\cite{AL1} for examples).


\subsection{Projective structure and topology}
\label{pro}
Although ${\rm Hom}\,[\P,G]$ is a very large space, it is a well defined limit ---
a projective limit ---
of a family of finite dimensional spaces. Each space of this so-called projective family is 
identified, as a manifold~\cite{AL1,AL2,AL3}, with some power of the Lie group $G$.
This projective structure is critically important. It gives us a good understanding
of  ${\rm Hom}\,[\P,G]$, allowing e.g.~to equip it with a rich variety of structures, 
from topology and 
measures~\cite{AL1,AL2} to differential calculus~\cite{AL3}. In a precise sense, the projective
structure reduces the task of dealing with a large infinite dimensional space to a 
problem in finite dimensions plus certain consistency conditions. Projective methods,
and their inductive "dual" counterparts, are therefore basic tools of the present approach, from its
foundations to current research.

We present next the projective structure of ${\rm Hom}\,[\P,G]$. In particular, this
structure gives rise to a natural topology on ${\rm Hom}\,[\P,G]$.
The compact space thus obtained is the space of generalized connections.
\subsubsection{The set of labels}
\label{L}
We start by introducing the directed set used as the set of labels of the projective family.
\begin{deff}
\label{def2}
A finite set $\{e_1,\ldots,e_n\}$ of edges is said to be independent if the
edges $e_i$ can intersect each other only at the points $s(e_i)$ or 
$r(e_i)$, $i=1,\ldots,n$.
\end{deff}
The edges in an independent set are, in particular, algebraically independent,
i.e.~it is not possible to produce identity paths by (non-trivial) 
compositions of the edges and their inverses. 

Let us consider subgroupoids of $\P$ 
generated by independent sets of edges $\{e_1,\ldots,e_n\}$. Recall that the 
subgroupoid generated by $\{e_1,\ldots,e_n\}$ is the smallest 
subgroupoid containing all the 
edges $e_i$, or explicitly, the subgroupoid whose objects are all the
points $s(e_i)$ and $r(e_i)$ and whose arrows are all possible
compositions of edges  $e_i$ and their inverses. Groupoids of this
type are freely generated, given the algebraic independence of the edges.
\begin{deff}
\label{def02}
The set of subgroupoids of $\P$ that are generated by independent sets of edges is called
the set $\L$ of tame subgroupoids.
\end{deff}
Clearly, the sets 
$\{e_1,\ldots,e_n\}$ and
$\bigl\{e_1^{\epsilon_1},\ldots,e_n^{\epsilon_n}\bigr\}$, where
$\epsilon_j=\pm 1$ (i.e.~$e_j^{\epsilon_j}=e_j$ or $e_j^{-1}$), generate
the same subgroupoid, and this is the only ambiguity in the choice of the
set of generators of a given groupoid $L\in\L$. Thus, a groupoid
$L\in\L$ is uniquely defined by a set $\bigl\{\s(e_1),\ldots,\s(e_n)\bigr\}$
of images corresponding to a set of independent edges. Notice that the union of the images 
$\s(e_i)$ is a graph in the manifold \Si, and therefore tame subgroupoids are
uniquely associated to analytic graphs.

Let us consider in the set $\L$ the partial-order relation defined by 
inclusion, i.e.~given $L,L'\in\L$, we will say that $L'\geq L$ if and only
if $L$ is a subgroupoid of $L'$. Recall that $L$ is said to be a subgroupoid 
of $L'$ if and only if all objects of $L$ are objects of $L'$ and for
any pair of objects $x,y$ of $L$ every arrow from $x$ to $y$ is an arrow
of $L'$. It is not difficult to see that $\L$ is a
directed set with respect to the latter partial-order, meaning that 
for any given
$L$ and $L'$ in $\L$ there exists $L''\in\L$ such that
$L''\geq L$ and $L''\geq L'$.
Let us remark that this is a point where analyticity is very important, allowing to show 
e.g.~that for every finitely generated subgroupoid $\Gamma\subset\P$ there is an
element $L\in\L$ such that~$\Gamma$ is a subgroupoid of $L$~\cite{AL1}.
\subsubsection{Projective family}
\label{AL}
We introduce the projective family, induce a compact Hausdorff topology on each of its members,
and show that consistent projections exist. Important results 
concerning the projective limit of such so-called compact Hausdorff
families were given in~\cite{AL2}.
\begin{deff}
\label{pf}
For every $L\in \L$,
let $\A_L:={\rm Hom}\,[L,G]$ denote the set of all morphisms from the subgroupoid
$L$ to the group $G$.
\end{deff}
Let $\{e_1,\ldots,e_n\}$ be a set of generators of $L\in\L$. 
Since the morphisms $L\to G$ are
uniquely determined by the images of the generators, one gets a
bijection $\rho_{e_1,\ldots,e_n}:\A_L\to G^n$, given by
\be
\label{bara4}
\A_L\ni\bar A\mapsto \bigl(\bar A(e_1),\ldots,\bar A(e_n)\bigr)\in G^n\,.
\ee
Thus, every set $\A_L$ is in 1-1 correspondence with some $G^n$.
Through this identification, every $\A_L$ becomes
a compact Hausdorff space. Notice that the 
topology induced
in $\A_L$ is independent of the choice of the generators, since maps of the form
\be
\label{bara5}
(g_1,\ldots,g_n)\mapsto \Bigl(g_{k_1}^{\epsilon_{k_1}},\ldots,
g_{k_n}^{\epsilon_{k_n}}\Bigr),
\ee
where $(k_1,\ldots,k_n)$ is a permutation of $(1,\ldots,n)$ and
$\epsilon_{k_i}=\pm 1$, are homeomorphisms $G^n\to G^n$. 

We will show next that the family of compact spaces
$\A_L,\ L\in\L$, is  a  compact Hausdorff projective family, meaning that given 
$L,L'\in\L$ such that~$L'\geq L$ there exists a
surjective and continuous projection $p_{L,L'}:\A_{L'}\rightarrow \A_L$
such that
\be
\label{grupg6}
p_{L,L''}=p_{L,L'}\circ p_{L',L''},\ \forall L''\geq L'\geq L\,.
\ee
For $L'\geq L$, the required projection 
\be
\label{pll}
p_{L,L'}:\A_{L'}\to \A_L 
\ee
is naturally defined to be the map
that sends each element of $\A_{L'}$ to its restriction to $L$. It is clear
that (\ref{grupg6}) is satisfied. 
Let us show that the maps $p_{L,L'}$ 
are surjective and continuous. Let $\{e_1,\ldots,e_n\}$ 
be generators of $L$ and $\{e'_1,\ldots,e'_m\}$ 
be generators of $L'\geq L$. Let us consider the decomposition of the edges
$e_i$ in terms of the edges $e'_j$:
\be
\label{bara1}
e_i=\prod_j ({e_{r_{ij}}'})^{\epsilon_{ij}},\ \ i=1,\ldots,n\, ,
\ee
where $r_{ij}$ and $\epsilon_{ij}$ take values in the sets
$\{1,\ldots,m\}$ and $\{1,-1\}$, respectively. An arbitrary element 
of $\A_L$ is
identified by the images $(h_1,\ldots,h_n)\in G^n$ of the ordered set
of generators $(e_1,\ldots,e_n)$. The map $p_{L,L'}$ will, therefore, be surjective
if and only if there are $(g_1,\ldots,g_m)\in G^m$ such that
\be
\label{bara2}
h_i=\prod_jg^{\epsilon_{ij}}_{r_{ij}},\ \ \forall i\,.
\ee
These conditions can indeed be satisfied, since they are independent.
In fact, since the edges $\{e_1,\ldots,e_n\}$ are independent, a given edge
$e'_k$ can appear at most once (in the form  $e'_k$ or $e'^{-1}_k$)
in the decomposition (\ref{bara1}) of a given $e_i$. 
As for continuity,
notice that, through the identifications (\ref{bara4}), the maps $p_{L,L'}$ 
correspond to projections $\pi_{n,m}:G^m\to G^n$,
\be
\label{bara3}
G^m\ni(g_1,\ldots,g_m)\stackrel{\pi_{n,m}}{\longmapsto}\Bigl({\textstyle
\prod_j}g^{\epsilon_{1j}}_{r_{1j}},\ldots,{\textstyle \prod _j}g^
{\epsilon_{nj}}_{r_{nj}}\Bigr)\in G^n\,,
\ee
which are continuous.
\subsubsection{Projective limit}
\label{projlim}
The general notion of  projective limit (see e.g~\cite{Ya}) applies in particular
to the current situation.
\begin{deff}
\label{pl}
The projective limit of the family $\{\A_L,p_{L,L'}\}_{L,L'\in\L}$ is the
subset $\A_{\infty}$ of the cartesian product $\times_{L\in\L}\A_L$ of those
elements $(A_L)_{L\in\L}$ that satisfy the consistency conditions 
\be
\label{eq1}
p_{L,L'}A_{L'}=A_L\ \ \ \ \forall \ L'\geq L\,.
\ee
\end{deff}
It is a simple, yet illustrative, exercise to show that ${\rm Hom}\,[\P,G]$ is in 1-1 
correspondence with the projective limit
$\A_{\infty}$. Let us  consider the map
\be
\label{phi}
\Phi: {\rm Hom}\,[\P,G]\to \A_{\infty}, \ \bar A\mapsto ({\bar A}_{|L})_{L\in\L} ,
\ee
where ${\bar A}_{|L}$ is the restriction of $\bar A$ to the tame subgroupoid $L$.
It is obvious that $({\bar A}_{|L})_{L\in\L}$ belongs to $\A_{\infty}$, 
i.e.~satisfies (\ref{eq1}). To prove that $\Phi$ is a bijection one just needs
to remember that every path $p\in\P$ belongs to some subgroupoid $L(p)\in\L$,
and therefore $\bar A(p)= {\bar A}_{|L}(p)$, $\forall L\geq L(p)$. If 
$\Phi(\bar A)=\Phi(\bar A')$, i.e.~if ${\bar A}_{|L}={\bar A'}_{|L}\ \forall L$,
we immediately get $\bar A(p)=\bar A'(p)\ \forall p$, i.e.~$\bar A=\bar A'$,
thus proving injectivity. Suppose now that we are given any $(A_L)_{L\in\L}\in\A_{\infty}$.
We construct its inverse image by $\bar A(p):=A_L(p)$, where $L$ is any tame 
subgroupoid such that $p\in L$. This is well defined, since given two such $L$
and $L'$ one can always find $L''\geq L,L'$, ensuring that 
$A_L(p)=p_{L,L''}A_{L''}(p)$ coincides with $A_{L'}(p)=p_{L',L''}A_{L''}(p)$.
As $\bar A$ thus defined is obviously a morphism, surjectivity is proven.
\begin{prop}
\label{prop0}
The map $\Phi$ (\ref{phi}) is a bijection.
\end{prop}
We henceforth identify ${\rm Hom}\,[\P,G]$ with the projective limit $\A_{\infty}$, by means of
the bijection $\Phi$ (\ref{phi}). In particular, the projections
\ba
p_L:\qquad\quad {\rm Hom}\,[\P,G] & \to & \A_L \nonumber \\
\label{eqa2}
\bar A & \mapsto & {\bar A}_{|L}\ ,L\in\L
\ea
correspond to the maps $\A_{\infty}\ni (A_L)_{L\in\L}\mapsto A_L\in\A_L$. It is clear that the 
following consistency conditions
\be
\label{consist}
p_L=p_{L,L'}\circ p_{L'},\ \forall L'\geq L,
\ee
corresponding to (\ref{eq1}), are satisfied.
\subsubsection{Natural topology}
\label{newtop}
Given the special nature of our particular projective family $\{\A_L\}_{L\in\L}$, 
two important results follow.
First, the projections $p_L$ above are guaranteed to be
surjective\footnote{A stronger result in fact holds, see lemma \ref{lem1}
in section \ref{classical}.}~\cite{AL1,AL2}. Second, the projective limit is naturally 
a compact Hausdorff space~\cite{MM,AL2}.
The projective limit topology in
${\rm Hom}\,[\P,G]$ is the weakest topology such that 
all the  projections $p_L$ (\ref{eqa2})  are continuous. 
We sketch below a simplified proof of the fact that the thus obtained topological space
is compact Hausdorff. We use arguments that are adapted from those given in~\cite{MM}.

Let us start by giving an equivalent description of the topology. 
Notice that the projections (\ref{eqa2})
are continuous if and only if the maps 
\be
\label{5}
{\rm Hom}\,[\P,G]\to G,\ \ \bar A\mapsto \bar A(e),
\ee
are continuous for every edge $e$, since every $\A_L$ is homeomorphic to
some $G^n$ and the topology on $G^n$ is generated by $G$-open sets. Since continuity
of all maps (\ref{5}) implies continuity of the maps
\be
\label{6}
\pi_p: {\rm Hom}\,[\P,G] \to G,\ \ \bar A\mapsto \bar A(p),
\ee
for all $p\in\P$, one can also characterize the topology on ${\rm Hom}\,[\P,G]$
as the 
weakest topology such that all maps $\pi_p$ (\ref{6}) are 
continuous.
Consider now the set of {\it all} maps from $\P$ to $G$, identified with the product space 
$\times_{p\in\P}G$. The product space is compact Hausdorff with the Tychonov topology. 
It is clear that the topology defined by the maps (\ref{6})   coincides  with the subspace topology 
on ${\rm Hom}\,[\P,G]$  as a subset of $\times_{p\in\P}G$.
The Hausdorff property of  ${\rm Hom}\,[\P,G]$ is inherited from $\times_{p\in\P}G$, and from the
fact that  ${\rm Hom}\,[\P,G]$ contains only  morphisms follows easily that it is closed,
therefore compact.
\begin{deff}
\label{def03}
The space ${\rm Hom}\,[\P,G]$, equipped with the weakest topology such that all
maps (\ref{eqa2}) (or equivalently (\ref{6})) are continuous, is called the 
space of generalized connections.
This compact Hausdorff space will hereafter be denoted by \ab.
\end{deff}

\subsection{Algebra of functions and inductive structure}
\label{inductive}
The projective characterization of the compact space \ab\
has as a  counterpart the inductive characterization of the corresponding
$C^{\star}$-algebra of continuous 
complex functions $C(\ab)$. 

Let us consider the family of $C^{\star}$-algebras
$\{C(\A_L)\}_{L\in\L}$, where $C(\A_L)$ is the algebra of continuous complex functions
on $\A_L$. The pull-back $p^*_{L,L'}$ of the projections $p_{L,L'}$ (\ref{pll})
define injective $C^{\star}$-morphisms
\be
\label{10}
 p^*_{L,L'}: C(\A_L)\to C(\A_{L'}),\ L'\geq L,
\ee 
that satisfy the consistency conditions, following from (\ref{grupg6}),
\be
\label{11}  
 p^*_{L,L''}=p^*_{L',L''}\circ p^*_{L,L'}, \ \forall L''\geq L'\geq L.
\ee
Such  a structure is called an inductive family. Turning to \ab, the pull-back
$p^*_L$ of the projections $p_L$ (\ref{eqa2}) 
define injective $C^{\star}$-morphisms
\be
\label{12}
p^*_{L}: C(\A_L)\to C(\ab)
\ee 
satisfying consistency conditions following from (\ref{consist}):
\be
\label{13}  
p^*_{L}=p^*_{L'}\circ p^*_{L,L'}, \ \forall L'\geq L.
\ee
Injectivity of $p^*_{L,L'}$ and $p^*_{L}$ follows from surjectivity
of $p_{L,L'}$ and $p_{L}$, respectively. Likewise, we see that $p^*_{L}$
and $p^*_{L,L'}$ are isometries, i.e.
\be
\label{iso}
\sup_{\bar \A}|p^*_L f|=\sup_{\A_{L'}}|p^*_{L,L'} f|=\sup_{\A_{L}}|f|,\ \forall 
f\in C(\A_L).
\ee
Let us consider the set  $\bigcup_{L\in\L}p^*_L C(\A_L)$  of all continuous
functions in \ab\ that are obtained by pull-back. It is obvious that this set is closed 
under complex conjugation. It also follows easily from (\ref{13}) (and the fact that
$\L$ is directed) that the above set is an algebra under pointwise multiplication.
It is therefore a $\star$-subalgebra of $C(\ab)$.
\begin{deff}
\label{cyl}
The $\star$-algebra ${\rm Cyl}(\ab):=\bigcup_{L\in\L}p^*_L C(\A_L)$ is called the 
algebra of cylindrical functions.
\end{deff}
Recalling the identification (\ref{bara4}) of every $\A_L$ with some $G^n$, it 
becomes clear that every cylindrical function, i.e.~every element of ${\rm Cyl}(\ab)$,
can be written in the form
\be
\label{8}
\ab\ni\Ab\mapsto f(\Ab)=F(\Ab(e_1),\ldots,\Ab(e_n)),
\ee
where $\{e_1,\ldots,e_n\}$ is a set of independent edges and $F:G^n\to\C$ is a continuous
function.
Notice that we can equally replace independent edges by arbitrary paths 
$\{p_1,\ldots,p_n\}$ in (\ref{8}), as paths can always be decomposed using independent
edges, and therefore $F(\Ab(p_1),\ldots,\Ab(p_n))$ can be written as 
$F'(\Ab(e_1),\ldots,\Ab(e_m))$, where $F'$ is again continuous.
\begin{prop}
\label{prop1}
The algebra ${\rm Cyl}(\ab)$ of cylindrical functions on \ab\ is dense on the
algebra $C(\ab)$ of continuous complex functions.
\end{prop}
This result follows from the Stone-Weierstrass theorem, since ${\rm Cyl}(\ab)$ 
is a $\star$-algebra, contains the identity function and clearly separates
points in \ab, as the functions $\ab\ni\Ab\mapsto\Ab(e)\in G$ separate points,
when all edges $e$ are taken into account.

This latter result, together with (\ref{10} -- \ref{13}), establishes that the
$C^{\star}$-algebra $C(\ab)$ is (isomorphic to) the so-called
inductive limit of the family of 
$C^{\star}$-algebras $\{C(\A_L)\}_{L\in\L}$ (see e.g.~\cite{WO}).

Let us mention that, besides ${\rm Cyl}(\ab)$, a whole (decreasing) sequence of 
$\star$-algebras ${\rm Cyl}^k(\ab)$, $k=\{1,2,\ldots\}\cup\{\infty\}$,
can be defined, exactly as ${\rm Cyl}(\ab)$, but considering only $C^k(\A_L)$ functions, 
i.e.~$C^k(G^n)$ functions $F$ in (\ref{8}). Proposition \ref{prop1} still holds for all
these subalgebras, as already $C^{\infty}$ functions separate points in $G$.
Differential calculus is naturally introduced in \ab\ by using derivations
of these algebras of differentiable functions. In particular, vector fields in \ab\ can be defined
by certain consistent families of vector fields in the finite dimensional spaces $\A_L$~\cite{AL3}. 

\subsection{Generalized gauge transformations and diffeomorphisms}
\label{gauge}
Two distinct groups act naturally and continuously on $\bar\A$. 
One of these is the group 
of natural transformations (see 
section \ref{cat}) of the set of functors ${\rm Hom}[\P,G]\equiv\ab$ . 
This group is well understood,
and is commonly accepted as the natural generalization of the group
$\G=C^{\infty}(\Sigma,G)$ of smooth local gauge transformations to the present
quantum context. The second group of interest is the group ${\rm Aut}(\P)$ 
of automorphisms
of the groupoid $\P$. It contains the group 
${\rm Diff}^{\omega}(\Sigma)$ of analytic diffeomorphisms as a subgroup.
Although extensions of ${\rm Diff}^{\omega}(\Sigma)$ are welcome, the group 
${\rm Aut}(\P)$ has not been studied yet. 

We begin by discussing the group of natural transformations of \ab.
In this case, natural transformations  
form the group, hereafter denoted by $\bar\G$, of all maps $\gg:\Sigma\to G$, under 
pointwise multiplication.  The action of $\bar \G$ on $\bar \A$ can be written as
$\bar \A\times\bar \G\ni(\Ab,\gg)\mapsto {\Ab}_{\gg}$ such that
\be
\label{3}
{\Ab}_{\gg}(p)=\gg(r(p))\Ab(p)\gg(s(p))^{-1}.
\ee
This action is readily seen to be continuous. In fact, since the topology on 
$\bar \A$ is the 
weakest  such that all maps $\pi_p$ (\ref{6}) are continuous, one can conclude
that a map
$\varphi:\ab\to\ab$ is continuous if and only if the maps $\pi_p\circ\varphi$
are continuous $\forall p$, which is obviously the case for elements of $\bar \G$.

Expression (\ref{3}) is a generalization of the action of smooth gauge 
transformations on the set of parallel transport functions $h_p(A)$ for smooth 
connections. It is therefore natural to accept $\bar \G$ as the generalized group
of gauge transformations on $\bar \A$. This extension of the gauge group is in fact 
required: since $\bar \A$ now contains arbitrary, e.g.~non continuous, morphisms
from $\P$ to $G$, to mod out only by smooth gauge transformations would 
leave spurious degrees of freedom untouched. This is seen 
most explicitly from the fact that \ab\ is actually homeomorphic to 
$\overline {\A/\G}\times\bar\G$~\cite{AL1,AL3,B1,MM,AL2,Ve}, where the gauge invariant space
$\overline {\A/\G}$ is the original space of generalized connections modulo gauge 
transformations~\cite{AI,AL1} (and $\bar\G$ is equiped with the Tychonov topology). A homeomorphism
relating $\overline {\A/\G}$ and the quotient space $\ab/{\bar\G}$ then follows, establishing the
equivalence between the current approach and the original manifestly gauge invariant one.

Let us now turn to  the group ${\rm Aut}(\P)$ of automorphisms
of the groupoid $\P$. By definition, an automorphism of $\P$ is an invertible functor
from $\P$ to itself. An element $F\in {\rm Aut}(\P)$ is therefore described
by a bijection of $\Sigma$ and a composition preserving bijection on the set of paths,
such that $F({\bf 1}_x)={\bf 1}_{F(x)}$, $\forall x\in\Sigma$. The action of 
${\rm Aut}(\P)$ on $\bar \A$ is given by
\be
\label{4}
\Ab\mapsto F\Ab\, :\  F\Ab(p)=\Ab(F^{-1}p),\ \forall p\in\P, \ F\in {\rm Aut}(\P).
\ee
The continuity of this action is clear, as $\pi_p\circ F=\pi_{F^{-1}p}$.
The group ${\rm Aut}(\P)$ contains as a subgroup the natural representation of
the group ${\rm Diff}^{\omega}(\Sigma)$ of analytic diffeomorphisms of $\Sigma$,
whose action on curves factors through the equivalence relation
that defines $\P$.

The group ${\rm Aut}(\P)$ therefore emerges, in the current quantum context, as the 
largest natural
extension of ${\rm Diff}^{\omega}(\Sigma)$. Extensions of
 ${\rm Diff}^{\omega}(\Sigma)$ are welcome. In fact, 
 the very formalism seems to require some sort of extension of 
${\rm Diff}^{\omega}(\Sigma)$, in the spirit of the extension $\bar \G$ of $\G$.
Going a little outside the scope of the present work, 
let us point out that the diffeomorphism 
invariant Hilbert space~\cite{ALMMT} one obtains from the kinematical Hilbert space $\H_0$
(see sections \ref{int} and \ref{m0}) 
is non-separable, when only ${\rm Diff}^{\omega}(\Sigma)$ is taken into 
account. (The invariant Hilbert space is, essentially, obtained by considering 
${\rm Diff}^{\omega}(\Sigma)$-orbits in $\H_0$, which are naturally realized as linear functionals
over ${\rm Cyl}^{\infty}(\ab)$.) Since a separable Hilbert space, 
although not absolutely required, is expected at this level, one is led to 
suspect that the group ${\rm Diff}^{\omega}(\Sigma)$ is too small.

One possible way out of this situation 
is to accept certain non-smooth transformations of $\Sigma$ as 
gauge~\cite{Z,T3,FR}.
For instance, 
it was shown
in~\cite{Z}  that the inclusion of piecewise analytic transformations is sufficient
to achieve separability. While these transformations are not fully motivated from the 
classical perspective, it appears that the quantum enlargement $\bar \A$ of $\A$
introduces additional spurious degrees of freedom that are no longer gauged away
by the action of classical smooth transformations. Furthermore, the replacement
of smooth transformations by a more general, perhaps combinatorial, group is not
unlikely to occur in the final theory of quantum gravity, as smoothness itself is
expected to dissolve at the Planck scale,
thus being a meaningful concept at the semi-classical regime only (see e.g.~\cite{T1,R2} and 
references therein for arguments along these lines).

The suggested requirement for a quantum enlargement of not only 
${\rm Diff}^{\omega}(\Sigma)$ but of ${\rm Diff}^{\infty}(\Sigma)$  itself also
removes some of the motivation for working with smooth curves and smooth 
diffeomorphisms, giving support to the idea that the differentiability class one
starts with might be irrelevant~\cite{Z,T1}. 

If these ideas turn out fruitful, then 
${\rm Aut}(\P)$ appears, in the current context, as the largest  natural group where to 
look for a quantum version of the diffeomorphism group.
A natural candidate is e.g.~the subgroup of ${\rm Aut}(\P)$ of those elements that are induced
by homeomorphisms of $\Sigma$.
 
It must be strongly stressed, however, that there are other options to deal with
the non-separability issue~\cite{T1,T3,T4}, and that possible extensions of the 
diffeomorphism group must be analyzed in depth and treated with great care, as they
potencially have a large impact on the quantum theory.

\section{Representations of the holonomy algebra}
\label{new3}
\subsection{$\bar \A$ as the spectrum of the  holonomy algebra}
\label{classical}
We have thus far shown that \ab\ is a  compact Hausdorff extension of the classical
configuration space $\A$, but its exact relation with $\A$ and how it allows a 
quantization of a 
particular algebra of classical configuration functions was not yet 
established. We will do that next, by showing that ${\rm Cyl}(\ab)$  is naturally isomorphic,
as a normed $\star$-algebra,  to the classical configuration algebra ${\rm Cyl}(\A)$
introduced in section \ref{int}. (This ultimately justifies the slight misuse of language 
in calling both ${\rm Cyl}(\A)$ and ${\rm Cyl}(\ab)$ the algebra of cylindrical functions.
Moreover, it is sometimes useful, and common practice, to actually identify  
${\rm Cyl}(\A)$ with ${\rm Cyl}(\ab)$, and the Ashtekar-Isham holonomy algebra with $C(\ab)$.)

Let us  define ${\rm Cyl}(\A)$ more precisely.
\begin{deff}
\label{cylA}
The algebra ${\rm Cyl}(\A)$ is the $\star$-algebra of all functions $f:\A\to\C$
of the form
\be
\label{9}
\A\ni A\mapsto f(A)=F(h_{e_1}(A),\ldots,h_{e_n}(A)),
\ee
where $\{e_1,\ldots,e_n\}$ is a set of independent edges and $F\in C(G^n)$.
${\rm Cyl}(\A)$ is a normed $\star$-algebra with respect to the supremum norm.
\end{deff}
It is clear that ${\rm Cyl}(\A)$ separates points in $\A$, again by the crucial fact,
mentioned in section \ref{ssi}, that holonomies separate points. Alternatively,
notice that ${\rm Cyl}(\A)$ can be described as the restriction of ${\rm Cyl}(\ab)$
to the faithful image of $\A\subset\ab$. The algebra ${\rm Cyl}(\A)$ is therefore
a viable set of classical configuration functions on which to base the quantization
process. (This is, of course, supplemented with the equally complete set
of momentum variables $E_{S,f}$, see section \ref{int}, so that coordinates in phase space 
can be defined.)
\begin{deff}
\label{HA}
The holonomy algebra ${\overline {\rm Cyl}}(\A)$ is the $C^{\star}$-completion 
of the normed
$\star$-algebra  ${\rm Cyl}(\A)$.
\end{deff}
Let us see that ${\overline {\rm Cyl}}(\A)$ is isomorphic to $C(\ab)$. 
This follows from denseness
of ${\rm Cyl}(\ab)$ (proposition \ref{prop1}) and the natural identification
between ${\rm Cyl}(\ab)$ and ${\rm Cyl}(\A)$.
The 1-1 correspondence between ${\rm Cyl}(\ab)$ and 
${\rm Cyl}(\A)$ is obvious. The isomorphism of normed algebras is ensured by
the following non-trivial fact~\cite{AL1}:
\begin{lem}
\label{lem1}
The maps
$$A\ni \A\mapsto (h(e_1,A),\ldots,h(e_n,A))\in G^n$$
are surjective, for every set $\{e_1,\ldots,e_n\}$  of independent edges.
\end{lem}
By lemma \ref{lem1}, the supremum of $|F(\Ab(e_1),\ldots,\Ab(e_n))|$ is already
attained in $\A\subset\ab$, and therefore the norm of  $F(\Ab(e_1),\ldots,\Ab(e_n))\in 
{\rm Cyl}(\ab)$ equals the norm of 
 $F(h(e_1,A),\ldots,h(e_n,A))\in {\rm Cyl}(\A)$. 
\begin{prop}
\label{CAisoHA}
The $C^{\star}$-algebra $C(\ab)$ is isomorphic to ${\overline {\rm Cyl}}(\A)$.
\end{prop}
In other words, \ab\ is the compact 
Hausdorff space, whose existence and uniqueness (up to homeomorphism) is guaranteed 
by the Gelfand-Naimark
theorem, on which the unital commutative $C^{\star}$-algebra 
${\overline {\rm Cyl}}(\A)$ is realized as an algebra of continuous functions.
In the $C^{\star}$-algebraic language, \ab\ is called the spectrum of 
${\overline {\rm Cyl}}(\A)$. It follows  from general topological arguments 
that \ab\ is a actually a compactification of $\A$, i.e.~the image of $\A\subset\ab$ is dense.

A quantization of the configuration
algebra ${\rm Cyl}(\A)$ is naturally defined to be a representation of the 
holonomy algebra. This is now seen to be the same as a representation of $C(\ab)$. 
The framework of representation theory of commutative
unital $C^{\star}$-algebras therefore comes into play, with measures on \ab\
(or, equivalently, states of the algebra $C(\ab)$) playing a crucial role.

\subsection{General structure of measures on \ab}
\label{mea}

Since \ab\ is a compact Hausdorff space, regular Borel measures are known to exist. 
In fact, normalized measures on \ab\ are in 1-1 correspondence with states of the
$C^{\star}$-algebra $C(\ab)$. (Recall that by a state it is meant a linear functional
$\omega: C(\ab)\to\C$ that is positive, i.e.~$\omega(ff^*)\geq 0$, $\forall f\in C(\ab)$,
and normalized, i.e.~$\|\omega\|=\om(1)=1$. Such linear functionals are necessarily 
continuous.) This follows from the Riez-Markov theorem:
\begin{theo}
\label{RM}
Let $X$ be a compact Hausdorff space. For any state \om\ of the $C^{\star}$-algebra
$C(X)$ there is a unique regular Borel probability measure $\m$ on $X$ such that
\be
\om(f)=\int_X f d\m,\ \forall f\in C(X).
\ee
\end{theo}
Every (regular Borel probability) measure $\m$ on \ab\ produces a cyclic \rep\ $\pi$ of 
$C(\ab)$, by multiplication operators on $L^2(\ab,\m)$:
\be
\label{cyc}
\left(\pi(f)\psi\right)(\bar A)=f(\bar A)\psi(\bar A),\ \psi\in L^2(\ab,\m),\ f\in C(\ab),
\ee
with cyclic vector $\Omega=1$ (i.e.~$\{\pi(f)\Omega,\ f\in C(\ab)\}$ is dense). Moreover, we know
from general $C^{\star}$-algebra results (see e.g.~\cite{BR}) that every cyclic \rep\ is
(unitarily equivalent to) a \rep\ of the above type (\ref{cyc}), and that general, non-cyclic,
\reps\ are obtained as direct sums of such cyclic \reps.

Measure theory on \ab\ is most conveniently addressed by combining the Riez-Markov
theorem with the projective-inductive structure.
\begin{deff}
\label{def3}
A family of measures $\{\m_L\}_{L\in\L}$, where $\m_L$ is a regular Borel probability 
measure on $\A_L$,
is said to be consistent if, $\forall L, L'$ such that $L'\geq L$, the measure
$\m_L$ coincides with the push-forward measure $(p_{L,L'})_*\m_{L'}$, i.e.~if
\be
\label{14}
\int_{\A_{L'}}p^*_{L,L'}f\, d\m_{L'}=\int_{\A_{L}}f\, d\m_{L}, \ \
\forall f\in C(\A_L).
\ee
\end{deff}
It is clear that a regular Borel probability measure $\m$ on \ab\ determines a 
consistent family $\{\m_L\}$ by
\be
\label{15}
\int_{\A_{L}}f\, d\m_{L}=\int_{\bar \A}p^*_{L}f\, d\m \ \
\forall f\in C(\A_L).
\ee
It turns out that the converse is also true~\cite{AL2}. To see this, let us consider a 
given consistent family $\{\m_L\}_{L\in\L}$. To simplify formulae, we work
with the corresponding family of states $\{\om_L\}_{L\in\L}$, where 
$\om_L(f)=\int_{\A_L} f d\m_L$. From $\{\om_L\}$ one can define a linear functional 
$\om:{\rm Cyl}(\ab)\to\C$ as follows. Let $F\in{\rm Cyl}(\ab)$. By definition, there exists
$L\in\L$ and $f\in C(\A_L)$ such that $F=p^*_L f$. The functional \om\ is defined by:
\be
\label{omega}
\om(F)=\om(p^*_Lf)=\om_L(f).
\ee
Let us check that this is a well defined functional. Let $F=p^*_{L'} f'$ be another way of 
writing the cylindrical function $F$. By considering $L''\geq L,L'$, one arrives at 
$F=p^*_{L''}p^*_{L,L''}f$ and $F=p^*_{L''}p^*_{L',L''}f'$, by  (\ref{13}). Since $p^*_{L''}$
is injective, we then get $p^*_{L,L''}f=p^*_{L',L''}f'$. The consistency property (\ref{14})
of the family now ensures that
\be
\label{ooo}
\om_{L'}(f')= \om_{L''}(p^*_{L',L''}f')=\om_{L''}(p^*_{L,L''}f)=\om_L(f).
\ee
The linear functional \om\ is clearly continuous, as
\be
\label{ocont}
|\om(p^*_Lf)|=|\om_L(f)|\leq\sup_{\A_L}|f|=\sup_{\bar \A}|p^*_Lf|,
\ee
by (\ref{iso}). Since ${\rm Cyl}(\ab)$ is dense, \om\ uniquely defines a continuous linear
functional $\om:C(\ab)\to \C$. Moreover, since $\|\om_L\|=\om(1)=1$ (and using again
$\sup_{\bar \A}|p^*_Lf|=\sup_{\A_L}|f|$), we obtain $\|\om\|=\om(1)=1$. Standard 
$C^{\star}$-algebra arguments show that the above conditions are sufficient for \om\ to 
qualify as a state of the algebra $C(\ab)$. Finally, using again the Riez-Markov theorem,
we obtain a measure satisfying (\ref{15}). Thus: 
\begin{prop}
\label{theo1}
Regular Borel probability measures on \ab\ are in 1-1 correspondence with consistent 
families of measures.
Explicitly, a consistent family $\{\m_L\}$ uniquely determines a measure $\m$
on \ab\ satisfying condition (\ref{15}).
\end{prop}
A measure on \ab\ is therefore equivalent to a consistent family of measures
on the finite dimensional spaces $\A_L$. We then reencounter a familiar situation
in quantum field theory, with the important difference that the 
finite dimensional configuration spaces $\A_L$ are now compact.

The projective-inductive structure is reflected in every Hilbert space $\H=L^2(\ab,\m)$.
Given a measure $\m$ with associated family $\{\m_L\}$, let
$\H_L$  denote the Hilbert space $L^2(\A_L,\m_L)$. It is clear that the pull-backs
$p^*_{L,L'}$ (\ref{10}) define injective inner product preserving maps 
$p^*_{L,L'}:\H_L \to \H_{L'}$, satisfying the
consistency conditions (\ref{11}). Likewise, the pull-backs
$p^*_L$ (\ref{12}) define linear transformations
$p^*_L:\H_L\to \H$, 
which are unitary when considered as maps 
\be
\label{z3}
p^*_L:\H_L\to p^*_{L}\H_{L}
\ee
onto their images, the closed subspaces $p^*_{L}\H_{L}$. Consistency conditions 
(\ref{13})
and the denseness of the subspace $\bigcup_{L\in\L} p^*_L\H_L\supset{\rm Cyl}(\ab)$  
characterize $\H$
as the so-called inductive limit of the inductive family of Hilbert spaces $\{\H_L\}_{L\in\L}$.
This structure is useful e.g.~in the construction of operators in $\H$, starting from consistent
families of operators in the spaces $\H_L$, and  finds application
in the quantization of  momentum, and more general, operators~\cite{AL3,T1}.

\subsection{The Ashtekar-Lewandowski representation}
\label{m0}
As discussed in the introduction, the Ashtekar-Lewandowski, or uniform, measure \mo\
is a central object in the canonical loop quantum gravity programme. Although several other
diffeomorphism invariant (and non-invariant) measures on \ab\ were constructed~\cite{B1,B2,AL2,AL3}, 
\mo\ is the only known
measure that supports a quantization of the flux variables $E_{S,f}$. The uniform measure
was introduced by 
Ashtekar and Lewandowski~\cite{AL1} in the $\overline {\A/\G}$  gauge invariant context,
and later on the original construction was generalized by Baez~\cite{B1} to produce the uniform
measure in \ab. In this section we present the uniform measure and briefly discuss
its most important properties.

The uniform   measure $\mo$  is constructed from  a consistent
family of measures determined only by the Haar measure $\mu_H$ on the group 
$G$. It can be defined as follows.
Let us consider the projective family $\{\A_L,p_{L,L'}\}_{L,L'\in\L}$. 
For $L\in\L$, let 
$(e_1,\ldots,e_n)$ be a ordered set of generators of the groupoid $L$,
and consider  the associated homeomorphism $\rho_{e_1,\ldots,e_n}:
\A_L\to G^n$ (\ref{bara4}). Let
$\m_H^n$ be the Haar measure on $G^n$ and let us denote by \mol\ the measure in
$\A_L$ that is obtained by push-forward of  $\m_H^n$ with respect to
$\rho^{-1}_{e_1,\ldots,e_n}$, i.e., $\mol:=(\rho^{-1}_{e_1,\ldots,e_n})_*
\,\m_H^n$, or
\be
\label{mol}
\int_{\A_L}fd\mol:=\int_{G^n}(\rho^{-1}_{e_1,\ldots,e_n})^*fd\m_H^n,\ 
\forall f\in C(\A_L).
\ee
By construction, \mol\ is a regular 
Borel probability measure on $\A_L$. Since $\m_H^n$ is a product of $n$ identical
measures, it is obvious that \mol\ is independent of the order of the generators
$\{e_1,\ldots,e_n\}$. Moreover, since the Haar measure  
$\m_H$ is invariant with respect to inversions $g\mapsto g^{-1}$,
$g\in G$, it follows that \mol\ is also independent of the choice of the set of 
generators of the groupoid  $L$. 

The family of measures
$\{\mol\}_{L\in\L}$ thus defined is consistent, also by the properties of the Haar
measure. In fact, given $L$ generated by $\{e_1,
\ldots,e_n\}$ and $L'$ generated by $\{e_1',\ldots,e_m'\}$, 
with $L'\geq L$, the consistency
conditions (\ref{14}) translate into 
\be
\m^n_H=(\pi_{n,m})_*\,\m^m_H\,,
\ee
or
\be
\label{meba8}
\int_{G^m}f(\pi_{n,m}(g_1,\ldots,g_m))d\m^m_H
=\int_{G^n}f(g_1,\ldots,g_n)d\m^n_H,\ \forall f\in C(G^n),
\ee
where $\pi_{n,m}:G^m\to G^n$ is the projection (\ref{bara3}).
It is not difficult to verify that the conditions
(\ref{meba8}) are satisfied, taking into account the invariance properties
of the Haar measure (see \cite{AL1} for a complete proof).

The family $\{\mol\}$ defines, by proposition \ref{theo1}, a regular Borel 
probability measure on $\bar\A$. This is the uniform measure \mo.

The simplest integrable functions on \ab\ are, of course, the cylindrical functions
\be
\label{111eq1}
f(\bar A)=F\bigl(\bar A(e_1),\ldots,\bar A(e_n)\bigr)\,,
\ee
with $\{e_1,\ldots,e_n\}$ a  set of independent edges and
$F\in C(G^n)$. For such functions we have simply:
\be
\label{111eq2}
\int_{\bar\A}f(\bar A)d\mo=\int_{G^n}F(g_1,\ldots,g_n)d\m^n_H.
\ee
Let us denote by $\H_0:=L^2(\ab,\mo)$ the Hilbert space defined by \mo. 
Notice that $\H_0$ is a non-separable space.
Notice also that an orthonormal basis 
of $\H_0$ is explicitly known. This is the important spin-network basis~\cite{RS,B3,T5}.
The Hilbert space $\H_0$
carries, of course, a cyclic \rep\ of  $C(\ab)$ by multiplication operators (\ref{cyc}).
We will refer to this \rep\ simply as the $\H_0$ \rep.

The most distinguished property of the $\H_0$ \rep\ is that it supports a 
quantization of the loop quantum gravity kinematical algebra, introduced in section \ref{int}.
Without going into any details (see e.g.~\cite{T1}), this goes as follows. Instead 
of the full holonomy algebra, let us consider its subalgebra ${\rm Cyl}^{\infty}(\A)$,
defined as ${\rm Cyl}^{\infty}(\ab)$ (section \ref{inductive}), and obviously isomorphic to it.
(This brings no loss of generality, as ${\rm Cyl}^{\infty}(\A)$ is dense. See moreover~\cite{OL}.)
It turns out that  the flux variables $E_{S,f}$ are naturally realized as derivations
$X_{S,f}$ of the algebra ${\rm Cyl}^{\infty}(\A)$, and that the Lie algebra -- let us call it
ACZ, after Ashtekar, Corichi and Zapata -- generated by ${\rm Cyl}^{\infty}(\A)$ and the set of
derivations $X_{S,f}$ is isomorphic to the Poisson algebra generated by the kinematical variables.
Actually, the Lie algebra ACZ is the rigorous way to define the kinematical Poisson algebra, 
as Poisson brackets among kinematical variables are {\it a priori} ill-defined, due to the 
particular smearing of connections and electric fields. It is only after proper regularization
that one obtains a well defined Lie algebra, and this is the Ashtekar-Corichi-Zapata 
algebra~\cite{ACZ}. Thus, a quantization of the kinematical algebra is defined to be
a (Dirac) \rep\ of the ACZ Lie algebra (meaning, of course, that we map real variables
to self-adjoint operators and that a factor $i$ is assumed in the commutators).   
It is clear that the assignments ${\rm Cyl}^{\infty}(\A)\ni f\mapsto f$ and 
$X_{S,f}\mapsto iX_{S,f}$ formally satisfy the commutation relations, in any Hilbert space
$L^2(\ab,\m)$, where $f$ is now seen as an element of ${\rm Cyl}^{\infty}(\ab)$, and
$iX_{S,f}$ as linear operators densely defined on ${\rm Cyl}^{\infty}(\ab)$. It turns out that
for the \mo\ measure, and only for that measure~\cite{T1,OL,ST}, the linear operators
$iX_{S,f}$ are actually self-adjoint. Thus, the $\H_0$ \rep\ extends to a quantization of
the kinematical algebra~\cite{ALMMT}. This \rep\ is  irreducible~\cite{ST1}, and it 
is the only known irreducible \rep\ of the loop quantum gravity kinematical algebra.

Let us see next that  the measure \mo\ is invariant under the induced action on \ab~(\ref{4})
of the group ${\rm Diff}^{\omega}(\Sigma)$ of analytic diffeomorphisms of $\Sigma$.
Notice that, by the Riez-Markov theorem and denseness of ${\rm Cyl}^{\infty}(\ab)$, it is sufficient
to check invariance on cylindrical functions. This is easily confirmed as follows. 
It is clear that for every analytic diffeomorphism  $\varphi\in{\rm Diff}^{\omega}(\Sigma)$ and every
set of independent edges $\{e_1,\ldots,e_n\}$, the diffeomorphic image
$\{\varphi e_1,\ldots,\varphi e_n\}$ is again 
an independent set of edges. In other words, for every 
tame subgroupoid $L\in\L$, its diffeomorphic image
$\varphi L$ is again a tame subgroupoid. Moreover, the spaces $\A_L$ and $\A_{\varphi L}$ are 
homeomorphic. Finally, both $\A_L$ and $\A_{\varphi L}$ are equiped with the same (Haar) measure,
leading to invariance. Explicitly, for every cylindrical function $f$ (\ref{111eq1}) we obtain 
from (\ref{111eq2}):
\ba
\nonumber
\int_{\bar\A}F(\varphi^{-1}\bar A(e_1),\ldots,\varphi^{-1}\bar A(e_n))d\mo
&=&\int_{\bar\A}F(\bar A(\varphi e_1),\ldots,\bar A(\varphi e_n))d\mo=\\ 
\label{18}
=\int_{G^n}F(g_1,\ldots,g_n)d\m^n_H
&=&\int_{\bar\A}F(\bar A(e_1),\ldots,\bar A(e_n))d\mo.
\ea
{}From invariance of the measure, a natural unitary
\rep\  $U_{D}$ of  ${\rm Diff}^{\omega}(\Sigma)$ in $\H_0$ follows:
\be
\label{19}
\left(U_{D}(\varphi)\psi\right)(\bar A)=\psi(\varphi^{-1}\bar A),\ 
\varphi\in{\rm Diff}^{\omega}(\Sigma),\psi\in\H_0.
\ee
We now turn to the  action (\ref{3}) of the (extended) gauge group $\bar \G$. Here we find an even
simpler situation, as $\bar \G$ acts within each space $\A_L$. Explicitly, for every $\gg\in\bar\G$
we have
\be
\label{ga}
\int_{\bar\A}F(\bar A_{\gg}(e_1),\ldots,\bar A_{\gg}(e_n))d\mo
=\int_{\bar\A}F(g_{11}\bar A(e_1)g_{12},\ldots,g_{n1}\bar A(e_n)g_{n2})d\mo,
\ee
where $g_{i1}:=\gg(r(e_i))$ and $g_{i2}:=\gg((s(e_i))^{-1}$ are fixed elements of $G$,
for each fixed set $\{e_1,\ldots,e_n\}$.  
{}From (\ref{111eq2}) and  invariance properties of the Haar measure $\m_H$ follows
that \mo\ is $\bar\G$-invariant. 
A representation of ${\bar \G}$ is therefore also obtained in $\H_0$.
In this case, the corresponding Gauss constraint
is immediately solved by the gauge invariant subspace of $\H_0$. This is a large
closed subspace, that can be obtained by closure of the well understood subspace
of gauge invariant cylindrical functions.
Unfortunately, this is not the case for the diffeomorphism constraint, as the only 
${\rm Diff}^{\omega}(\Sigma)$-invariant elements of $\H_0$ are the constant 
functions~\cite{B4,MTV}. As already mentioned, the solution space of the diffeomorphism
constraint lies in the algebraic dual of the space  ${\rm Cyl}^{\infty}(\ab)$~\cite{ALMMT}.



\section*{Acknowledgements}
\noindent I 
thank Professor Gennadi Sardanashvily for the invitation to contribute to the special issue of
{\it International Journal
of Geometric Methods in Modern Physics} on "Advanced Geometric Techniques in 
Gauge Theory".
I am most
grateful to Jos\'e Mour\~ao  for numerous discussions and suggestions.
I also thank Thomas Thiemann for useful discussions.  
This work was supported in part by 
POCTI/33943/MAT/2000, CERN/P/FIS/43171/2001, POCTI/FNU/\-49529/2002 and POCTI/FP/FNU/50226/2003.






\end{document}